\documentstyle[preprint,prc,eqsecnum,aps,epsf]{revtex}

\tightenlines   

\begin{document}
\draft

\def\lsim{\thinspace{\hbox to 8pt{\raise -5pt\hbox{$\sim$}\hss{$<$}}}\thinspace}
\def\rsim{\thinspace{\hbox to 8pt{\raise -5pt\hbox{$\sim$}\hss{$>$}}}\thinspace}

\title{Two-Body T-Matrices without Angular Momentum Decomposition:
Energy and Momentum Dependencies}

\author{ Ch.~Elster, J.H.~Thomas}
\address{
Institute of Nuclear and Particle Physics,  and
Department of Physics, \\ Ohio University, Athens, OH 45701}
    
\author{ W. Gl\"ockle}
\address{
 Institute for Theoretical Physics II, Ruhr-University Bochum,
D-44780 Bochum, Germany.}

\vspace{10mm}

\date{\today}

\maketitle

\begin{abstract}   
The two-body t-matrix is calculated directly as function of
two vector momenta for different Malfliet-Tjon type potentials.
At a few hundred MeV projectile energy the total amplitude is
quite a smooth function showing only a strong peak in forward direction.
In contrast 
the corresponding partial wave contributions, whose number
increases with increasing energy, 
become more and more oscillatory with increasing energy. 
The angular and momentum dependence of the full amplitude
is studied and displayed on as well as off the energy shell 
 as function of positive and negative energies. The behavior of the
t-matrix in the vicinity of 
bound state poles and  resonance poles in the second energy
sheet is studied. It is found that the angular dependence of T
exhibits a very characteristic behavior in the vicinity of those
poles, which is given by the Legendre function corresponding
to the quantum number either of the bound state or the resonance (or virtual)
state. This behavior is illustrated with numerical examples. 

\end{abstract}

\vspace{10mm}


\pagebreak

 \narrowtext 


\section{Introduction}

\hspace*{10mm}
At low energies in the MeV and the few tenth of MeV region very few
angular momenta contribute to the nucleon-nucleon (NN) scattering 
process. Consequently a description using angular momentum decomposition
is an adequate tool for carrying out scattering calculations.
However, at intermediate energies, i.e. energies of a few hundred MeV, and
higher energies very many angular momenta contribute to the scattering 
amplitude. There those individual contributions to the scattering 
amplitude at a fixed high angular momentum oscillate strongly in angle,
whereas the total amplitude is much smoother.
This suggests the direct determination of $T$ as function of the initial
and final momentum vectors avoiding angular momentum decomposition
totally. For NN scattering investigations of this kind have already
been undertaken \cite{holz,rice,hiltrop,huber}.

The choice of momentum vectors as adequate variables is also suggested
from the NN force. The dependence on momentum vectors in the case of
the widely used one-boson-exchange force is for instance rather simple,
whereas the partial wave representation of this force leads to complicated
expressions \cite{bonn}. This is already apparent in the most simple case
of a scalar meson propagator, $1/(({\bf q'}-{\bf q})^2-\mu^2)$, 
which is the central ingredient to any NN force. In a partial wave 
decomposition this is represented in the form $\frac {1}{q'q} Q_l(z)$,
where $z=({\bf q'}^2+{\bf q}^2 + \mu^2)/2q'q$ and $Q_l(z)$ is the
Legendre function of the second kind. For large values of $l$ the latter
requires some care in order to be handled numerically correctly.

Scattering of more than two particles requires two-body t-matrices
off-the-energy-shell (off-shell for short notation) as dynamical input,
which are apparently easier to handle if they enter the calculations
as smooth functions instead of strongly varying partial wave components.
Of course those remarks also apply to the treatment of scattering processes
of more than two particles at intermediate energies, which themselves
are also treated more economically and transparently using momentum
vectors instead of partial wave representations. Calculations of 
three and more particles use as input fully off-shell two-body t-matrices,
whose properties as functions of arbitrary initial and final momenta 
and in general positive and negative energies should be well understood.
Specifically at negative energies there may be bound state poles 
and in the second energy sheet there may be poles related to virtual states
and resonances.

Our aim in this article is to generate two-body t-matrices directly 
in a three-dimensional form and display their properties as  function
of the magnitudes of the off-shell momenta, the angle between the two
momentum vectors and of the energy. We are not aware of a similar study
in this generality in the literature. Usually only partial wave projected
amplitudes are displayed and discussed \cite{redish}. In Section II
we describe our solution of the two-body Lippmann-Schwinger equation
directly as function of the momentum vectors and illustrate the on- and
off-shell properties of the resulting t-matrices obtained with
simple Yukawa type two-nucleon potentials. In Section III we discuss the 
pole structure of the t-matrix as function of the energy and illustrate
its angular and energy behavior at and around bound state poles.
In Section IV we discuss and illustrate the behavior of the t-matrix
for virtual and resonant states in the second energy sheet. We conclude
in Section V.

\section{The On- and Off-Shell Two-Body T-Matrix at Positive Energies}

\hspace*{10mm}
Two-body scattering is governed by the Lippmann-Schwinger equation
\begin{equation}
 T = V + V G_0 T, \label{eq:2.1}
\end{equation} 
where $V$ is the two-body (e.g. two-nucleon) potential, $G_0=(z-H_0)^{-1}$
the free two-body propagator and $T$ the transition operator.
In momentum space its matrix elements $T({\bf q'},{\bf q},z) \equiv
\langle {\bf q'}|T(z)|{\bf q}\rangle$ obey the integral equation
\begin{equation}
T({\bf q'},{\bf q},z)=V({\bf q'},{\bf q}) + \int d^3 q''
V({\bf q'},{\bf q''})\frac {1}{z-\frac {q''^2}{m}}  
T({\bf q''},{\bf q},z).   \label{eq:2.2}
\end{equation} 
Here ${\bf q}$ are the relative momenta, $m$ the common mass of the
two particles and $z$ an arbitrary energy. 
We use a nonrelativistic framework. 
In this article we restrict ourselves to two spinless
particles and local potentials.
Therefore, $V({\bf q'},{\bf q})$ as well as $T({\bf q'},{\bf q},z)$ are
scalar functions:
\begin{equation}
V({\bf q'},{\bf q})= V(q',q,{\bf \hat q}'\cdot {\bf \hat q}) \label{eq:2.3}
\end{equation}
and 
\begin{equation}
T({\bf q'},{\bf q})= T(q',q,{\bf \hat q}'\cdot {\bf \hat q}). \label{eq:2.4}
\end{equation}
In Eq.~(\ref{eq:2.4}) we dropped the parametric dependence on the energy
$z$. This notation then leads to the explicit form of Eq.~(\ref{eq:2.2})
\begin{equation}
T(q',q,x')= V(q',q,x') + \int_0^{\infty} dq'' q''^2 \int_{-1}^{1} dx''
    \int_0^{2\pi} d\varphi'' V(q',q'',y) \frac{1}{z-\frac{q''^2}{m}}
          T(q'',q,x''), \label{eq:2.5}
\end{equation}
where $x'={\bf \hat q}'\cdot {\bf \hat q}$, $x''={\bf \hat q}''\cdot 
{\bf \hat q}$, 
and $y={\bf \hat q}''\cdot {\bf \hat q}'$. We can express $y$ through $x'$ and
$x''$ as
\begin{equation}
y=x'x'' +\sqrt{1-x'^2} \sqrt{1-x''^2} \cos{\varphi''}  \label{eq:2.6}
\end{equation}
where the arbitrary azimuthal angle $\varphi$ for ${\bf \hat q}$ is chosen to
be zero. If we define
\begin{equation}
v(q',q,x',x)\equiv \int_0^{\infty} d\varphi V(q',q,x'x+\sqrt{1-x'^2}
  \sqrt{1-x^2}\cos \varphi),  \label{eq:2.7}
\end{equation}
the integral equation Eq.~(\ref{eq:2.5}) becomes
\begin{equation}
T(q',q,x')= \frac{1}{2\pi} v(q',q,x',1) +\int_0^{\infty} dq'' q''^2
\int_{-1}^{1} dx'' v(q',q'',x',x'') \frac{1}{z-\frac{q''^2}{m}}
 T(q'',q,x'') \label{eq:2.8}
\end{equation}
This is a two-dimensional integral equation in the off-shell momenta 
$q'$ ($q''$) and the cosine of the `scattering angle' $x'$ ($x''$).

\hspace*{10mm}
In this Section we consider the solutions of the integral equation 
Eq.~(\ref{eq:2.8}) at positive energies, i.e. we choose $z\equiv 
E+i\varepsilon=\frac{q_0^2}{m} +i\varepsilon$, corresponding to the
incoming momentum ${\bf q_0}$. In order to obtain insight into the
behavior of the t-matrix, we will consider the on-shell element
$T(q_0,q_0,x,E)$, whose square is proportional to the differential cross
section as well as the half-shell, $T(q,q_0,x,E)$, and fully off-shell,
$T(q,q',x,E)$ t-matrix. 

\hspace*{10mm}
We solve the two-dimensional integral equation
typically using 24 or 32 q-points and 24 x-points. The Cauchy
Singularity is separated into a principal value part and a
$\delta$-function part, and the principal value singularity is treated
by subtraction. The integration interval for the q-integration is covered
by mapping the Gauss-Legendre points $u$ from the interval (0,1) via 
$q=c \tan{(\frac{\pi}{2} u)}$ to the interval (0,$\infty$). Typical
values of c are 1000 MeV/c. 

\hspace*{10mm}
A very stringent test for our numerics
is the off-shell unitarity relation, which is a direct consequence of
Eq.~(\ref{eq:2.2}). In our two-dimensional form it reads
\begin{equation}
Im T(q',q,x') = -\frac {\pi}{2} m q_0 \int_{-1}^1 dx'' \int_0^{2\pi}
d\varphi'' T(q',q_0,y)T^{\star}(q,q_0,x'') \label{eq:2.9}
\end{equation}
where $y$ is given in Eq.~(\ref{eq:2.6}). With Eq.~(\ref{eq:2.9}) we
allowed for the most general case of the unitarity relation, where the
energy $z=E=\frac{q_0^2}{m}$ is not related to the incoming momentum
$q$, thus $q\neq q' \neq q_0$. In our numerical tests
Eq.~(\ref{eq:2.9}) was fulfilled for arbitrary $q',q$, and $x'$ values
with an accuracy below 0.001\% with the above quoted number of
integration points.

As main application we choose potentials of the Malfliet-Tjon \cite{MT}
type, i.e.
\begin{equation}
V(r) = V_r \frac{e^{-\mu_R}}{r} - V_A \frac{e^{-\mu_A}}{r},
\label{eq:2.10}
\end{equation}
and consequently
\begin{equation}
V({\bf q'},{\bf q})= \frac{1}{2\pi^2}\left( \frac{V_R}{({\bf q'}-
      {\bf q})^2 + \mu_R^2} - \frac{V_A}{({\bf q'}-{\bf q})^2 + \mu_A^2}
 \right).  \label{eq:2.11}
\end{equation}
In the case of a Malfliet-Tjon type potential the $\varphi$-integration
of Eq.~(\ref{eq:2.7}) can be carried out analytically with the result
\begin{eqnarray}
v(q',q,x',x)&=&\frac {1}{\pi} \left [ \frac {V_R} 
 {\sqrt{(q'^2+q^2-2q'qx'x +\mu_R)^2 -4q'^2q^2(1-x'^2)(1-x^2)}} \right .  \\
\nonumber
& &\left . - \frac {V_A} {\sqrt{(q'^2+q^2-2q'qx'x +\mu_A)^2
-4q'^2q^2(1-x'^2)(1-x^2)}} \right]. \label{eq:2.12}
\end{eqnarray}
The parameters used for V are given as $V^{(I)}$ in Table I. Note that
they are slightly different from the ones used in Ref.~\cite{MT}.

\hspace*{10mm}
As first numerical example we would like to demonstrate the connection of
the angle depended on-shell amplitude $T(q_0,q_0,x,E)$ and its
representation in terms of partial wave amplitudes,
\begin{equation}
T(q_0,q_0,x) = \sum_{l=0}^{\infty} \frac{2l+1}{4\pi} T_l(q_0) P_l(x),
 \label{eq:2.13}
\end{equation}
where $T_l(q_0)=\frac{2}{\pi} \frac{1}{q_0 m} e^{i\delta_l(q_0)} 
 \sin{\delta_l(q_0)}$. The quantity $\delta_l(q_0)$ is the phase-shift
for a given angular momentum $l$ and is determined in a standard manner.
In Fig.~1 we show  $Re\; T(q_0,q_0,x,E)$ at 300 and 800~MeV laboratory
energies together with partial wave sums up to a given angular momentum
$l$. Note that $E=E_{lab}/2$. The strong peak of $Re\; T(q_0,q_0,x,E)$ 
in forward direction requires
high orders of Legendre polynomials for a correct description. This is
of course especially pronounced for the higher energy. In Fig.~2 we
display $Im\; T(q_0,q_0,x,E)$ at the same energies together with its
representation in partial wave sums. It can be seen that $Im\;
T(q_0,q_0,x,E)$ needs less partial wave amplitudes for its correct
representation, the reason being that $Im\; T_l$ is proportional to $\sin
^2 \delta_l$,whereas $Re\; T_l$ is proportional to $\cos \delta_l \sin
\delta_l$. For large values of $l$ the phase shifts become small, thus
$Im\; T_l$ decreases with $\delta_l^2$, whereas $Re\; T_l$ only decreases
proportional to $\delta_l$.

\hspace*{10mm} 
An overview over the angular dependence of the full on-shell amplitude 
$T(q_0,q_0,x,E)$ as function of the energy is given in Fig.~3. Starting
from a relatively flat angular distribution at lower energies the peaking
in forward direction develops with increasing energy. At the same time
the angular range where the cross section is flat and small increases
with increasing energy, indicating that forward scattering dominates at
higher energies.

\hspace*{10mm}
Next we consider the half-shell amplitude $T(q,q_0,x,E)$ for two
energies. Fig.~4 shows that $T(q,q_0,x,E)$ is rather small and
structureless for all off-shell momenta $q$, with the exception of
$q$ being close to the on-shell momentum $q_0$. The most
general amplitude, the fully off-shell amplitude $T(q,q',x,E)$ is
displayed in Figs.~5 and 6 (the real part is shown) 
as function of one off-shell momentum 
$q$ and the angle $x$ for two fixed momenta $q'$. Contrary
to what one might expect, the strongest forward peaking does not occur
for $q$ being close to on-shell, but for $q=q'$. This agrees with the
behavior of the driving term, which peaks for $q=q'$. We found this
behavior for all energies $E>0$.

\hspace*{10mm}
All numerical and graphical examples considered so far refer to a
potential of Malfliet-Tjon type with repulsive and attractive parts 
(potential $V^{(I)}$ in Table~I). Its strength is such that it supports
a bound state at $E=$-2.23~MeV. Though this potential is of quite
simple character, it captures essential features of NN interaction
models based on meson exchange with respect to the propagator structure.

\section{The Off-Shell T-Matrix at Negative Energies}

\hspace*{10mm}
In the context of Faddeev-Yakubovsky equations two-body t-matrices
need to be evaluated at negative energies. In a three-body system
the energy argument for the two-body t-matrix is given as
$E=E_{tot}-\frac {3}{4m}q^2$ \cite{Wbook}. Here $E_{tot}$ is the total
energy of the three particle system and $\frac {3}{4m} q^2$ is the
kinetic energy of the relative motion of the third particle with respect
to the interacting pair, which is described by the t-matrix. For an 
interacting three-body system the relative momentum $|{\bf q}|$ is
not conserved. Therefore it can have arbitrary values and $E$ covers all
energies below $E_{tot}$. Thus we are interested to see, whether the
angular dependence of the t-matrix evaluated at negative energies is
similar to the one observed at positive energies.  A second consideration
is that bound states of the two-body system lead to poles in the t-matrix.
The angular dependence at and around a pole should be dictated by the
one of the bound state. We will investigate these questions and provide
numerical illustrations.

\hspace*{10mm}
The formal solution to the Lippmann-Schwinger equation, Eq.~(\ref{eq:2.1}),
is given by
\begin{equation}
 T(z) = V + V \frac {1}{z-H} V, \label{eq:3.1}
\end{equation} 
where $H$ is the full two-body Hamiltonian. If this Hamiltonian supports
a bound state $|\phi_b \rangle$ at $z=E_b$, it follows immediately that
\begin{equation}
 T(z) \;\; \stackrel{z \rightarrow E_b}{\longrightarrow} \;\;
  V|\phi_b \rangle \frac {1}{z-E_b}\langle \phi_b|V . \label{eq:3.2}
\end{equation} 

\noindent
In momentum space representation Eq.~(\ref{eq:3.2}) reads
\begin{equation}
 T({\bf q'},{\bf q},z) \;\; \stackrel{z \rightarrow E_b}{\longrightarrow}
 \;\;  \langle {\bf q'}|V|\phi_b \rangle \frac {1}{z-E_b}\langle 
\phi_b|V|{\bf q} \rangle . \label{eq:3.3}
\end{equation} 
The bound state obeys $H|\phi_b \rangle = E_b |\phi_b \rangle$ and has 
a certain fixed angular momentum l, such that
\begin{equation}
\langle {\bf q}|\phi_b\rangle = \phi_{b,l}(q) Y_{lm}({\bf \hat q}). 
\label{eq:3.4}
\end{equation} 

Since T is a scalar quantity, its behavior at and around the pole has to
have the form
\begin{eqnarray}
T({\bf q'},{\bf q},z) & \stackrel{z \rightarrow E_b}{\longrightarrow}&
\sum_m Y_{lm}({\bf \hat q'}) g_l(q') \frac {1}{z-E_b} Y_{lm}^{\star}
({\bf \hat q}) 
g_l(q) \\ \nonumber
 & & =\frac {2l+1}{4\pi} P_l({\bf \hat q'} \cdot {\bf \hat q}) 
\frac {g_l(q') g_l(q)}
{z-E_b}  \\ \nonumber
& & \equiv \frac {R_l(q',q,{\bf \hat q'} \cdot {\bf \hat q})}{z-E_b}.
 \label{eq:3.5}
\end{eqnarray}
Here
\begin{equation}
g_l(q) = \int_0^{\infty} dq' q'^2 v_l(q,q') \phi_{b,l}(q')
\label{eq:3.6}
\end{equation}
with
\begin{equation}
v_l(q,q')= \frac {2}{\pi} \int_0^{\infty} dr r^2 j_l(qr) V(r) j_l(q'r).
    \label{eq:3.7}
\end{equation}

\hspace*{10mm}
From Eq.~(\ref{eq:3.5}) it can be clearly seen that the angular
dependence of T exhibits a very characteristic behavior in the vicinity
of the bound state poles, which is given by the Legendre function
corresponding to the angular quantum number of the bound state.
In order to illustrate the pole behavior, we choose the potential 
$V^{(II)}$ of Table I, which supports a s-wave bound state at 
$E_s$= -190.16~MeV and a p-wave bound state at $E_p$= -14.629~MeV.
These binding energies are determined in a standard manner solving the
Schr\"odinger equation for a fixed angular momentum. This is a simple,
one-dimensional problem, whose solution also provides the function
$g_l(q)$ and thus the residue 
$R_l(q',q,{\bf \hat q}' \cdot {\bf \hat q})$
from Eq.~(\ref{eq:3.5}). The values of the binding energies can also be
obtained by solving the two-dimensional integral equation
Eq.~(\ref{eq:2.8}) and determining the pole position from the solution.
Choosing the same integration points $q$ in the partial wave projected,
one-dimensional form and the two-dimensional form, the bound state
energies $E_b$ agree very well with the pole positions $E_{pol}$. For
example, for 40 q-points (and 32 x-points) we find $E_b(l=0)$=
-190.162~MeV which has to be compared to $E_{pol}(l=0)$=
-190.164~MeV. Similarly, we find $E_b(l=1)$= -14.6296~MeV compared
to $E_{pol}(l=1)$= -14.6296~MeV. These results can be pushed to higher
accuracy if desired. We also determine the residues at each pole from
the solution of the two-dimensional integral equation and illustrate our
result in Table~II for the arbitrary choice of $q'=q=q_0=\sqrt{m|E|}$
and the angle averaged quantity $\bar T_l \equiv \frac {1}{c_l} 
\int_{-1}^1 dx P_l(x) T(q_0,q_0,x,E) (E-E_{pol}(l))$. 
As demonstrated in Table~II, we
approach the poles from both sides and the numbers closest to the poles
agree very well with the corresponding residues 
calculated directly from the partial wave projected problem.

\hspace*{10mm}
The angular dependence of $ T(q_0,q_0,x,E)$ with $q_0=\sqrt{m|E|}$ for
energies close to the two poles is displayed in Fig.~7. Both figures
show that near and at the pole the t-matrix shows the characteristic
behavior of the Legendre function associated with the angular momentum
quantum number of the corresponding bound state. In Fig.~8 we show the
angular dependence of $T(q_0,q_0,x,E)$ in the whole energy range around
and in between the bound state poles. In order to include both poles we
show $(E-E_s)(E-E_p) T(q_0,q_0,x,E)$ in Figs.~7 and 8. Starting from
very small values of $|E|$, the angular dependence is first of s-wave
character, then turns to a p-wave shape at and near the p-wave pole and
then develops the forward peak known from corresponding positive
energies. Note that due to the multiplicative factors, the peak turns
upward in the two figures. When $|E|$ reaches the s-wave pole, $T$ turns
back to the pure s-wave behavior and then finally flips back into a
strong forward peak.

\hspace*{10mm}
The fact that the angular dependence at negative energies is reminiscent
of that at the corresponding positive energies, except for the
characteristic behavior near the poles, leads us to suspect that
the real parts of $T$ might be quite similar to each other at energies
of equal magnitude. This turns out to be the case as demonstrated in
Fig.~9, where we display $ Re \; T(q_0,q_0,x,E)$ for the potential $V^{(I)}$
for different values $|E|$. It should be noted that the equality of 
$Re \; T(q_0,q_0,x,E)$ for positive and negative energies is not the trivial
consequence of $Re T \approx V$, which does not hold. In order to
demonstrate that $V$ is significantly different from $Re T$, we also
display $V$ in Fig.~9. Comparing $Re\; T(q_0,q_0,x,E)$ at the different
energies, we have to conclude that the rescattering terms, $Re T -V$,
behave more and more similar to each other for the same absolute values
of the energy.

\hspace*{10mm}
Finally in Fig.~10 we display the real part of the fully off-shell 
t-matrix $Re \; T(q,q',x,E)$ 
as function of $q$ and $x$ for fixed energy
$E=200$~MeV and fixed momenta $q'=250$~MeV/c and $q'=1000$~MeV/c.
As in Figs.~5 and 6 for positive energies $Re \; T$ is  most strongly peaked
at $q=q'$, which can  be expected once the information of Figs.~5
and 6 is known.

\section{The Off-Shell T-Matrix in the Second Energy Sheet}

\hspace*{10mm}
Two-body t-matrices might exhibit resonant behavior at positive energies
or show a strong energy dependence near $E=0$ due to a virtual state.
This latter case is realized for instance in the NN system for the
partial wave state $^1$S$_0$. Our goal is to locate those resonances in
the second energy sheet and investigate the characteristic angular
dependence connected with the resonance or virtual state.

\hspace*{10mm}
The transition to the second energy sheet requires an analytic
continuation of the Lippmann-Schwinger equation, Eq.~(\ref{eq:2.2}),
into the second energy sheet, which we briefly describe here \cite{Wbook}.
For a complex energy $z$ located in the upper half plane, we modify the
integration path as indicated in Fig.~11. The contribution along the
closed path II gives a residue and we find
\begin{eqnarray}
T({\bf q'},{\bf q},z) &=& V({\bf q'},{\bf q}) - i\pi m q_z \int d{\hat q}''
 V({\bf q'},{\hat q}''q_z) T({\bf \hat q}''q_z, {\bf q},z)  \nonumber \\
 & & + \int_{\cap} d^3 q'' V({\bf q'},{\bf q''})\frac {1}{z-\frac
{q''^2}{m}} T({\bf q''},{\bf q},z).   \label{eq:4.1}
\end{eqnarray}

\noindent
Here $q_z = \sqrt{mz}$, and the symbol at the second integral indicates
the deformed integration path I.
Since we deformed the integration path such that it is located
above the energy $z$, we are able to take  $z$
into the lower half of the complex plane without hitting a singularity
in the propagator. Once the energy $z$ is located in the lower half
plane, we can return with the integration path I to the real axis and
have instead of Eq.~(\ref{eq:4.1})
\begin{eqnarray}
T({\bf q'},{\bf q},z) &=& V({\bf q'},{\bf q}) - i\pi m q_z \int d{\hat q}''
V({\bf q'},{\bf {\hat q}}''q_z) T({\bf {\hat q}}''q_z, {\bf q},z)  \nonumber \\
& & + \int  d^3 q'' V({\bf q'},{\bf q''})\frac {1}{z-\frac{q''^2}{m}}
T({\bf q''},{\bf q},z).   \label{eq:4.2}
\end{eqnarray}
This equation is valid on the second energy sheet, which is reached from
the upper rim of the cut along the positive real energy axis in the
physical sheet. Due to the additional imaginary term,
Eq.~(\ref{eq:4.2}) has to be supplemented by another equation, which we
obtain by choosing ${\bf q}'={\bf {\hat q}}q_z$,
\begin{eqnarray}
T({\bf {\hat q}}'q_z, {\bf q},z) &=& 
V({\bf {\hat q}}''q_z,{\bf q}) - i\pi m q_z
  \int d{\hat q}'' V({\bf {\hat q}}'q_z, {\bf {\hat q}}''q_z)T({\bf
{\hat q}}''q_z,{\bf q},z)
 \nonumber  \nonumber\\ 
 & & + \int  d^3 q'' V({\bf {\hat q}}'q_z,{\bf q''}) 
\frac {1}{z-\frac{q''^2}{m}}
  T({\bf q''},{\bf q},z).   \label{eq:4.3}
\end{eqnarray}

\hspace*{10mm}
Similar to the bound state, which induces a nontrivial solution for the
homogeneous equation related to Eq.~(\ref{eq:2.1}), the homogeneous set
of equations related to Eqs.~(\ref{eq:4.2}) and (\ref{eq:4.3}) has 
a nontrivial solution at discrete values of $z$. These discrete values
either correspond to resonances with $Re\; z > 0$ and $Im\; z < 0$ or to
virtual states with $Re\; z < 0$ and $Im\; z = 0$. The fact that this
homogeneous set of equations has a nontrivial solution together with the
compactness property of the integral kernel means that $T(z)$ has a pole
at those energies $z$. We are interested not only in determining the
positions of those poles but also in understanding the residues and
their angular dependence.

\hspace*{10mm}
As it is obvious from Eq.~(\ref{eq:4.3}), the driving term singles out
the first entry as complex number. In order to simplify the formal steps
leading to a determination of the residue, it is convenient to
supplement the set of equations given in Eqs.~(\ref{eq:4.2}) and
(\ref{eq:4.3}) by another set in which the driving term has a complex
entry in the second argument. The two sets can then be combined using
the following matrix notation

\begin{eqnarray}
\left(\matrix{T({\bf q'},{\bf q},z)\:\:\: T({\bf q}', {\bf {\hat q}}'q_z,z) \cr
   T({\bf {\hat q}}'q_z, {\bf q},z)\:\:\: T({\bf {\hat q}}'q_z,{\bf
{\hat q}}q_z,z) \cr}\right)
&=&\left(\matrix{ V({\bf q'},{\bf q})\:\:\: V({\bf q}', {\bf {\hat q}}'q_z) \cr
       V({\bf {\hat q}}'q_z, {\bf q})\:\:\: V({\bf {\hat q}}'q_z,{\bf
{\hat q}}q_z) \cr}\right)
 \\ \nonumber
 & & \\ \nonumber 
&+& \left(\matrix{\int d^3 q'' V({\bf q'},{\bf q}'')\:\:\: 
       q_z^3 \int d{\hat q}'' V({\bf q'},{\bf {\hat q}}''q_z) \cr
  \int d^3 q'' V({\bf {\hat q}}'q_z,{\bf q}'')\:\:\: q_z^3 \int d{\hat q}''
          V({\bf {\hat q}}'q_z,{\bf {\hat q}}''q_z) \cr}\right) \\ \nonumber
 & & \\ \nonumber
&  \times &
 \left(\matrix{\frac{1}{z-\frac{q''^2}{m}}\:\:\: 0 \cr
            0\:\:\: -i\pi \frac{m}{q_z^2} \cr}\right) 
\left(\matrix{ T({\bf q}'',{\bf q},z)\:\:\: 
  T({\bf q}'', {\bf {\hat q}}q_z,z) \cr
   T({\bf {\hat q}}''q_z, {\bf q},z)\:\:\: T({\bf {\hat q}}''q_z,{\bf
{\hat q}}q_z,z) \cr}\right).
 \label{eq:4.4}
\end{eqnarray}

\noindent
Introducing the appropriate matrices, we write Eq.~(4.4) as 
\begin{equation}
\tilde{\bf T} =\tilde{\bf V}+\tilde{\bf V} \tilde{\bf G} \tilde{\bf T}.
    \label{eq:4.5}
\end{equation}

\hspace*{10mm}
We also need to study the corresponding homogeneous problem, which we
want to write in the following form
\begin{equation}
\lambda(z) \tilde{\bf \chi} = \tilde{\bf V}\tilde{\bf G}\tilde{\bf \chi}.
\label{eq:4.6}
\end{equation}
In this form the eigenvalue is $\lambda(z)$ and the energy $z$ is a 
parameter. Since $\tilde{\bf V}\tilde{\bf G}$ is a compact
operator, there is a discrete set of eigenvalues, which accumulate at 
$\lambda(z)=0$ \cite{weinberg}. The physical resonances occur at those
values $z\equiv E_{res}$, for which $\lambda(E_{res})=1$.

\noindent
In the following we choose $z=E_{res}$. Then we have
\begin{equation}
\tilde{\bf \chi}=\tilde{\bf V}\tilde{\bf G}\tilde{\bf \chi}.
\label{eq:4.7}
\end{equation}
Since the kernel is nonsymmetric, we also have to consider the left hand
eigenvalue problem
\begin{equation}
\tilde{\bf \Theta}^T = \tilde{\bf \Theta}^T \tilde{\bf V} \tilde{\bf G}.
 \label{eq:4.8}
\end{equation}
Defining
\begin{equation}
\tilde{\bf \Phi}^T \equiv \tilde{\bf \Theta}^T \tilde{\bf V}
\label{eq:4.9}
\end{equation}
we deduce
\begin{equation}
\tilde{\bf \Phi}^T = \tilde{\bf \Phi}^T  \tilde{\bf G} \tilde{\bf V}
\label{eq:4.10}
\end{equation}
or
\begin{equation}
\tilde{\bf \Phi}= \tilde{\bf V}^T \tilde{\bf G} \tilde{\bf \Phi}.
\label{eq:4.11}
\end{equation}
Since $\tilde{\bf V}^T = \tilde{\bf V}$, we obtain $\tilde{\bf \Phi}=
\tilde{\bf \chi}$.

\hspace*{10mm}
In the immediate neighborhood of $z=E_{res}$ and as a consequence of
\begin{equation}
\tilde{\bf T}(z) = (1 - \tilde{\bf V} \tilde{\bf G})^{-1} \tilde{\bf V}
\label{eq:4.12}
\end{equation}
one has
\begin{equation}
\tilde{\bf T}(z)\;\; \stackrel{z \rightarrow E_{res}}{\longrightarrow} \;\;
\tilde{\bf \chi} (1-\lambda(z))^{-1} \frac{1}{N} \tilde{\bf \Theta}^T
  \tilde{\bf V}, 
\label{eq:4.13}
\end{equation}
where $N$ is a normalization factor. 
In the neighborhood of $z=E_{res}$ we can put
\begin{equation}
\lambda(z) \approx 1 + \lambda'(z)\mid_{z=E_{res}} (z-E_{res})
\label{eq:4.14}
\end{equation}
and obtain
\begin{equation}
\tilde{\bf T}(z)\;\; \stackrel{z \rightarrow E_{res}}{\longrightarrow}
\;\; \tilde{\bf \chi} \frac{1}{z-E_{res}}
\frac{-1}{\lambda'(z)\mid_{z=E_{res}}} \frac{1}{N} \tilde{\bf \chi}^T.
\label{eq:4.15}
\end{equation}
For the case of a bound state pole it is easy to prove that 
$\frac{-1}{\lambda'(z)\mid_{z=E_{res}}} \frac{1}{N} =1$ for a
normalized bound state $|\phi_b\rangle$ and $|\chi\rangle =
V|\phi_b\rangle$. If we consider the right hand side of
Eq.~(\ref{eq:4.15}) as function of an auxiliary strength factor to the
potential, we can adopt the normalization of the bound state and define
\begin{equation}
\tilde{\bf T}(z)\;\; \stackrel{z \rightarrow E_{res}}{\longrightarrow}
\;\; \tilde{\bf \chi} \frac{1}{z-E_{res}} \tilde{\bf \chi}^T.
\label{eq:4.16}
\end{equation}

\hspace*{10mm}
The final remark concerns the scalar nature of $\tilde{\bf T}({\bf q'},
{\bf q},z)$. Since a resonant state has a unique angular momentum $l$,
the function $\chi({\bf q})$ will have the form
\begin{equation}
\chi({\bf q})=  \chi_l(q) Y_{lm}({\bf \hat q}), \label{eq:4.17}
\end{equation}
and we have to conclude that
\begin{equation}
\tilde{\bf T}({\bf q'},{\bf q},z) \;\; \stackrel{z \rightarrow
E_{res}}{\longrightarrow} \;\;  \frac{2l+1}{4\pi} \chi_l(q')
 \frac {1}{z-E_{res}} \chi_l(q) P_l({\bf \hat q}'\cdot {\bf \hat q}).
\label{eq:4.18}
\end{equation}
If $Re E_p >0$ and if $Im E_p <0$ sufficiently small, the the t-matrix
will feel the nearby pole also for real, positive energies $z$, and a
resonance will occur in the differential cross section. In the case of a
virtual state, like for $^1$S$_0$ in NN scattering, the pole is located at 
$Re E_p <0$ and $Im E_p =0$. For sufficiently small values of $|Re E_p|$
the t-matrix will be strongly enhanced near and at $z=0$.

\hspace*{10mm}
For our numerical realization we rewrite Eqs.~(\ref{eq:4.2}) and
(\ref{eq:4.3}) analogously to Eq.~(\ref{eq:2.8}) as
\begin{eqnarray}
T(q',q,x')&=& \frac {1}{2\pi} v(q',q,x,1)  \nonumber \\
  & & + \int_0^{\infty} dq'' q''^2 \int_{-1}^1 dx'' v(q',q'',x',x'')
   \frac {1}{z-\frac{q''^2}{m}} T(q'',q,x'')  \nonumber \\
 & & - i \pi m q_z \int_{-1}^1 dx'' v(q',q_z,x',x'') T(q_z,q,x'') 
 \label{eq:4.19}
\end{eqnarray}
and
\begin{eqnarray}
T(q_z,q,x')&=& \frac {1}{2\pi} v(q_z,q,x,1)  \nonumber \\
 & & + \int_0^{\infty} dq'' q''^2 \int_{-1}^1 dx'' v(q_z,q'',x',x'')
 \frac {1}{z-\frac{q''^2}{m}} T(q'',q,x'') \nonumber \\
  & & - i \pi m q_z \int_{-1}^1 dx'' v(q_z,q_z,x',x'') T(q_z,q,x'').
\label{eq:4.20}
\end{eqnarray}
Here $z \equiv E=|E| e^{i\phi}$ with $\phi <0$ and $q_z= \sqrt{mE}
=\sqrt{m|E|} e^{i\phi/2}$. 
The nontrivial solution to the homogeneous system of Eq.~(\ref{eq:4.7})
has a fixed angular momentum. When employing Eq.~(\ref{eq:4.17}), we 
obtain 
\begin{equation}
\chi_l(q) =  \int_0^{\infty} dq' q'^2 v_l(q,q')\frac{1}{z-\frac{q'^2}{m}}
\chi_l(q') -i\pi mq_z v_l(q,q_z) \chi_l(q_z)  \label{eq:4.21}
\end{equation}
and
\begin{equation}
\chi_l(q_z)=  \int_0^{\infty} dq' q'^2 v_l(q_z,q')
\frac{1}{z-\frac{q'^2}{m}}\chi_l(q') -i\pi mq_z v_l(q_z,q_z) \chi_l(q_z).
\label{eq:4.22}
\end{equation}

We used the above given equations Eqs.~(\ref{eq:4.21}) and 
(\ref{eq:4.22}) to determine the location of the resonances in the second 
energy sheet. For s-waves they are usually called virtual states and are
located on the negative energy axis. For partial waves with $l=1$ or 
higher, the energy eigenvalues have a positive real part and a negative
imaginary part. For varying potential strength, they move along trajectories
in the complex energy plane. In the following, we numerically study two 
different cases, namely a s-wave virtual state supported by potential
model $V^{(V)}$ of Table~I and a p-wave resonance of potential model
$V^{(IV)}$ of Table~I. The corresponding trajectories are listed in
Table~III and shown in Fig.~12.

\hspace*{10mm}
For the case of the s-wave virtual state we started from potential model
$V^{(V)}$ of Table~I. The s-wave phase shift of this potential has 
an effective range expansion with a scattering length $a_s = -23.5818$~fm
and an effective range $r_s = 2.8789$~fm. Using these values, the pole 
position for the S-matrix can be estimated via the effective range
expansion as
\begin{equation}
q^a_v = i \left [ \frac {1}{r_s} - \sqrt{\frac {2}{r_s|a_s|} +
  \frac {1}{r_s^2}} \right ] , \label{eq:4.23}
\end{equation}
which leads in our specific case to a position of the virtual state
$E_{q_v}^a = - \frac {(q^a_v)^2}{m}  = -0.06738$~MeV, where we used
$m = 938.9$~MeV. This number is very close to the exactly calculated
value given as $E_p(l=0) = -0.06663$~MeV, which is given in Table~III.
In the same table we quantify the s-wave trajectory as a function of the
strength parameter $V_A$ of potential $V^{(V)}$. 

For obtaining a p-wave resonance state, we start from potential
$V^{(III)}$ of Table~I, which supports a p-wave bound state, and 
decrease the attraction by decreasing the strength parameter $V_A$
until the bound state turns into a resonance state. Selected values for the
so obtained trajectory for the p-wave resonance are listed in 
Table~III. Of course, this model does not correspond to the reality of
an NN force, even for the lowest value of $V_A=4.7$ given in Table~III,
the binding energy of the s-wave bound state is still $E_s = -52.52$~MeV.
Nevertheless, this example illustrates in a simple manner, what can be 
expected for other cases like an effective nucleon-nucleus interaction,
which supports low energy resonances for certain angular momentum states.
Qualitatively the same picture would emerge.

\hspace*{10mm}
We would like to mention that we solved the homogeneous set of
Eqs.~(\ref{eq:4.21}) and (\ref{eq:4.22}) by the very efficient 
power method \cite{MT,Wbook}. Regarding the notation of Eq.~(\ref{eq:4.7}),
one has to determine the eigenvalue $\lambda(z)$ and vary the energy
such that $\lambda(z)=1$. For the potentials used here, the largest 
eigenvalue in magnitude was always an unphysical one with a negative 
real part generated by the repulsive short range piece of the force.
Once the largest eigenvalue is determined, we introduce a new integral
kernel, consisting of the old one minus that specific eigenvalue. The
in this way defined new kernel has then the physical eigenvalue as the
largest one in magnitude.

\hspace*{10mm}
Next we investigate the solution of the t-matrix in the second energy 
sheet  as given by Eqs.~(\ref{eq:4.19}) and (4.20).
We are interested in verifying the location of the pole as well as the
angular dependence of the residue at the pole as given in
Eq.~(\ref{eq:4.18}). We illustrate our findings for the potential
$V^{(IV)}$ of Table~I, which has a p-wave resonant state at
$E_{res}(l=1)=(4.3177 -i 2.2386)$~MeV and for the potential $V^{(V)}$ of
Table~I, which has a virtual s-wave state at $E_{res}(l=0)=
-0.06663$~MeV. For the p-wave resonance we show in Figs.~13 and 14
the angular dependence of $Re\;[ (E-E_{res}(l=1))T(q_E,q_0,x,E)]$ as function of 
the complex energy $E$ located along two straight lines going through
the pole position. In Fig.~13 we start on the real axis at $E=4.3$~MeV and
successively increase the imaginary part of $E$. For $E=(4.3 -i 2.4)$~MeV
we clearly see an angular dependence characteristic of a p-wave residue.
Since the width of the resonance is relatively small, the p-wave
behavior is present along the whole vertical energy line including the
point on the real axis. For our second choice of energy line (Fig.~14),
a declined line starting from zero energy, starts with a behavior being a
mixture of s- and p-wave, but still relatively flat. Approaching the
resonance, the shape becomes predominantly the one of a $P_1(x)$ given
by the p-wave residue. 

\hspace*{10mm}
A corresponding study based on the potential
model $V^{(V)}$ is shown in Fig.~15 for the negative real axis in the
second sheet, where the neighborhood of the virtual pole position
is considered. In the
vicinity of the virtual state, the residue exhibits perfect s-wave 
characteristics.

\hspace*{10mm}
Finally, we would like to demonstrate the structure of the t-matrix as
given in Eq.~(\ref{eq:4.18}) in a numerical example. First, we
numerically verify that $\lim_{z\rightarrow E_{res}}(z -E_{res})
T(q',q,x,z)$ behaves like $P_1(z)$. Then we determine $\chi_1(q)$ by
comparing the numerically calculated quantity $(z -E_{res})T(q',q,x,z)$
to the form given in Eq.~(\ref{eq:4.18}). Instead of dividing by
$P_1(x)$ we use
\begin{equation}
\bar{\chi}_1(q)= \lim_{z\rightarrow E_{res}} \sqrt{2\pi \int_{-1}^1
dx P_1(x) (z -E_{res})T(q,q,x,z)}. \label{eq:4.24}
\end{equation}
The values of $\bar{\chi}_1(q)$ for z approaching $E_{res}$ are shown for a
few 
arbitrarily selected momentum points $q$ in Table~IV. They stabilize
for $z$ approaching the pole position $E_{res}=(4.318 - i 2.239)$~MeV.
The last row in Table~IV shows the values obtained from the solution
of the homogeneous set of Eqs.~(\ref{eq:4.21}) and (\ref{eq:4.22}).
They have been normalized at one $q$ point to the $\chi_1(q)$ extracted
from $T(q,q,x,z\rightarrow E_{res})$. The agreement is perfect.

\hspace*{10mm}
Finally, we directly verified the separable structure of
$T(q',q,x,z\rightarrow E_{res})$ as given in Eq.~(\ref{eq:4.18}) by 
evaluating  $\tilde T= 2\pi \int_{-1}^1 dx P_1(x) (E-E_{res})
T(q',q,x,E)$ very close to $E=E_{res}$ for different 
$q'\neq q$ and comparing them to the values obtained via Table~IV.
The  agreement is again perfect and the values for a selected set of
momentum points $q'$ and $q$ are given  in  Table~V.

\section{Summary}

\hspace*{10mm}
Two nucleon scattering at intermediate energies of a few hundred MeV
requires quite a few angular momentum states in order to achieve convergence
of e.g. scattering observables. This is even more true for the scattering
of three or more nucleons upon each other. An alternative approach to the
conventional one, which is based on angular momentum decomposition, is to
work directly with momentum vectors, specifically with the magnitudes of the
momenta and the angles between them. We formulated and numerically
illustrated this alternative approach for the case of two-body scattering,
which includes the approach towards 
bound states and resonances.  The angular dependence
of the two-body t-matrix is directly determined from the Lippmann-Schwinger
equation, which now is a two-dimensional integral equation in contrast
to the one-dimensional one for a fixed angular momentum 
in a partial wave formulation. This two-dimensional integral equation
is quite easily numerically tractable. We determined the angular
dependence of the on-shell, half-shell and fully off-shell t-matrix 
as function of the scattering energy and different choices of momenta.
As two-body force we concentrated on a superposition of an attractive and
repulsive Yukawa interaction, which is typical for nuclear physics.
We neglected spin degrees of freedom in all our studies.

\hspace*{10mm}
We want to briefly summarize our results. The on-shell 
t-matrix develops a 
strong forward peak as the energy increases, which is more and more
difficult to build up in a calculation based on angular momentum
decomposition, but relatively simple accessible in our approach using
momentum vectors. The angular dependence of a half-shell t-matrix
is strong only around the on-shell momentum and rather mild otherwise.
For a fully off-shell t-matrix $T(q,q',x,E)$ the strong angular 
dependence occurs for $q=q'$, which do not necessarily have to coincide
with the on-shell momentum. At negative energies the t-matrix has
poles located at the bound state energies, if those exist. 
As example we investigated
s- and p-wave bound states. The numerically determined t-matrix turned
out to be very well under control even quite close to the the bound
state poles, where the homogeneous version of the Lippmann-Schwinger
equation has a nontrivial solution. We determined the angular dependence
of the t-matrix at the two poles, at energies between them and at
energies way below the deepest bound state. Directly at the poles the
angular behavior displays the characteristics of the Legendre polynomial
of the same angular momentum $l$ as the bound state. Between the poles
as well as for energies below the last bound state the t-matrix
exhibits the same forward peaking as visible at positive energies.
This latter result is interesting by itself. The angular dependence 
at positive and negative energies is very similar. More quantitatively,
we found that the real parts of the t-matrix are extremely close to each
other at positive and negative energies of equal magnitude long before
this statement becomes trivial due to the validity of the Born 
approximation $T=V$. 

\hspace*{10mm}
Finally we studied the analytical continuation of the Lippmann-Schwinger
equation into the second energy sheet, which is reached through the
cut along the positive real axis of the physical sheet. In the lower
half plane possible resonance poles of the t-matrix are located, which
are of course of interest only if they are close to the real axis.
As example we studied a p-wave resonance and mapped out its pole 
trajectory by varying the potential strength. At the pole, the t-matrix
achieves a separable form, which we verified numerically. We also
verified the characteristic angular dependence of the t-matrix close
to the resonance, which corresponds to the Legendre polynomial of degree
$l$ of the angular momentum state of the resonance. For negative
energies in the second sheet we investigated the pole trajectory of
a virtual s-wave state, which is of interest in the NN system for the
quantum number $^1$S$_0$. 

\hspace*{10mm}
Summarizing we can state that the two-dimensional Lippmann-Schwinger
equation can be handled quite easily in a numerically very reliable
manner. In this approach one determines directly the angular dependence
of the t-matrix for arbitrary momenta and energies.  Once supplemented
by spin degrees of freedom this approach will be of interest in the NN
system. In addition, this approach will be generalizable to systems
with three particles, like three nucleons or two nucleons and a meson.
In the case of three nucleons,  Faddeev calculations at e.g. 150~MeV and
higher are getting quite tedious because of the very many orbital angular
momentum states involved \cite{ndprep} and a direct, three-dimensional 
approach 
appears to be preferable. First steps in this direction are under way.

\vfill
\acknowledgements
This work was performed in part under the auspices of the
U.~S.  Department of Energy under contract No. DE-FG02-93ER40756 with
Ohio University and the NATO Collaborative Research Grant 960892.
We thank the Ohio
Supercomputer Center (OSC) for the use of their facilities
under Grant No.~PHS206. The authors would like to thank S.P.~Weppner
and X.D.~Zhang for their help in preparing some of the figures.



\pagebreak

\begin{table}
\caption{ Parameters of the Malfliet-Tjon type potentials.
As conversion factor we use units such that $\hbar c$=197.3286 MeVfm=1.}

\begin{tabular}{|c|cccc|}
\mbox{  } & $V_A$ & $\mu_A$ [MeV] & $V_R$  & $\mu_R$ [MeV] \\ \tableline
\tableline
$V^{(I)}$ & 3.1769  & 305.86  & 7.291  & 613.69 \\ 
$V^{(II)}$ &   -    &    -    &  10.0  & 2000.  \\
$V^{(III)}$ & 6.0   & 305.86  & 7.291  & 613.69 \\  
$V^{(IV)}$ & 5.1   & 305.86  & 7.291  & 613.69 \\  
$V^{(V)}$ & 2.6047   & 305.86  & 7.291  & 613.69 \\  
\end{tabular}
\end{table}

\begin{table}
\caption{ Determination of $\bar T_l \equiv 1/c_l \int_{-1}^1 dx
P_l(x) T(q_0,q_0,x,E) (E-E_{pol}(l))$ as function of $E$ close to the
$l=1$  and $l=0$ poles.The values for the constants are 
$c_0$=2 and $c_1$=2/3. The entry p.w. indicates the value determined
from the partial wave projected problem.}
\begin{tabular}{|cc||cc|}
$E$ [MeV]  & $\hat T_1(E)$ [MeV fm$^2$] & $E$ [MeV] &   $\hat T_0(E)$
[MeV fm$^2$] \\ \tableline \tableline

-14.60 & 1.08424 & -189.8 & 4.37911  \\
-14.61 & 1.08513 & -189.9 & 4.37847 \\
-14.62 & 1.08597 & -190.0 & 4.37772  \\
-14.63 & 1.08588 & -190.1 & 4.37722 \\
-14.64 & 1.08778 & -190.2 & 4.37693 \\
-14.65 & 1.08857 & -190.3 & 4.37608 \\ \tableline
p.w.  & 1.08684 &  p.w. & 4.37685 \\
\end{tabular}
\end{table}

\begin{table}
\caption{ Pole trajectories in the complex energy plane as function
of the strength parameter $V_A$ for the potentials $V^{(III)}$ (p-wave
pole trajectory) and $V^{(V)}$ (s-wave virtual state trajectory).}
\begin{tabular}{|cc||cc|}
$ V_A$ ($V^{(III)}$) & $E_{res}(l=1)$ [MeV] & $ V_A$ ($V^{(V)}$) &
 $E_{res}(l=0)$ [MeV] \\ \tableline \tableline
5.4 &    0.9267 -i 0.1821 & 2.6047 & -0.06663 \\
5.3 &    2.1972 -i 0.7101 & 2.6   &  -0.07380 \\
5.2 &    3.3254 -i 1.4134 & 2.5   &  -0.31004 \\
5.1 &    4.3177 -i 2.2386 & 2.4   &  -0.69870\\
5.0 &    5.1801 -i 3.1552 & 2.3   &  -1.22970 \\
4.9 &    5.9178 -i 4.1413 & 2.2   &  -1.89209  \\
4.8 &    6.5358 -i 5.1806 & 2.1   &  -2.67878 \\
4.7 &    7.0388 -i 6.2597 & 2.0   &  -3.57733 \\
\end{tabular}
\end{table}
 
\begin{table}
\caption{ Determination of $\bar{\chi}_1(q)$ in the vicinity of
the p-wave resonance of potential model $V^{(IV)}$ as function of
the complex energy E.}
\begin{tabular}{|c|cccc|}
E [MeV] & $\bar{\chi}_1$(q=94.89 MeV/c) & $\bar{\chi}_1$(q=292.73 MeV/c) &
  $\bar{\chi}_1$(q=527.50 MeV/c) & $\bar{\chi}_1$(q=758.09 MeV/c) \\
  \tableline \tableline
4.32 -i 2.0  & 1.105 -i 0.400 & 1.856 -i 0.729 & 1.149 -i 0.473 &
                                                 0.491 -i 0.212 \\
4.32 -i 2.10 & 1.103 -i 0.400 & 1.851 -i 0.729 & 1.144 -i 0.471 &
                                                 0.488 -i 0.207 \\
4.32 -i 2.20 & 1.101 -i 0.400 & 1.846 -i 0.729 & 1.139 -i 0.468 &
                                                 0.484 -i 0.202 \\
4.32 -i 2.23 & 1.101 -i 0.400 & 1.844 -i 0.729 & 1.138 -i 0.467 &
                                          0.483 -i 0.201 \\ 
4.32 -i 2.4  & 1.098 -i 0.400 & 1.835 -i 0.729 & 1.129 -i 0.461 &
                                          0.477 -i 0.192 \\
4.32 -i 2.5  & 1.096 -i 0.400 & 1.830 -i 0.729 & 1.124 -i 0.459 &
                                          0.473 -i 0.187 \\
 \tableline
 p.w.        & 1.102 -i 0.400 & 1.844 -i 0.729 & 1.138 -i 0.468 &
                                          0.482 -i 0.200 
      
\end{tabular}
\end{table}
 
\begin{table}
\caption{Comparison of $\bar T = 2\pi \int_{-1}^1 dx P_1(x)(E-E_{res})
T(q,q',x,E)$ with $\bar{\chi}_1(q)
\bar{\chi}_1(q')$ for a fixed value $q'$ and different values
of $q$ at the p-wave resonance of potential model $V^{(IV)}$.
The energy E for calculating $\bar T$ was fixed at $E=(4.32 -i
2.23)$~MeV.}
\begin{tabular}{|cc|cc|}
 q [MeV/c] & q' [MeV/c] & $\bar T [MeV fm^2]$ & 
 $\bar{\chi}_1(q)\bar{\chi}_1(q') [MeV fm^2]$ \\ \tableline \tableline

292.73  & 94.89 & 1.7398 -i 1.5395  & 1.7397 -i 1.5398 \\
527.50  & 94.89   & 1.0668 -i 0.9683  & 1.0667 -i 0.9689 \\
758.09  & 94.89   & 0.4522 -i 0.4134  & 0.4521 -i 0.4140 \\ 
\end{tabular}
\end{table}

\pagebreak

\noindent
\begin{figure}
\caption{Angular dependence for the real part of the on-shell t-matrix,
 $Re\; T(q_0,q_0,x,E)$. 
At $E_{lab}=$ 300~MeV (a) the dashed line represents
the partial wave sum up to $j=2$, the dash-dotted line the sum up to
$j=4$ and the solid line the sum up to $j=6$. The solid bullets stand for
the 3D calculation. At $E_{lab}=$ 800~MeV (b) the dashed line represents the
partial wave sum up to $j=6$, the dash-dotted line stands for the sum up to 
$j=9$, and the solid line for the sum up to $j=12$. Again, the solid
bullets represent the 3D calculation.  \label{fig1}}
\end{figure}

\noindent
\begin{figure}
\caption{Angular dependence for the imaginary part of the on-shell t-matrix, 
$Im T(q_0,q_0,x,E)$. At $E_{lab}=$ 300~MeV (a) the dashed line represents
the partial wave sum up to $j=2$, the solid line the sum up to $j=4$.
The solid bullets stand for the 3D calculation. At $E_{lab}=$ 800~MeV (b)
the dashed line represents the partial wave sum up to $j=3$, the solid
line the sum up to $j=6$. The bullets stand for the 3D calculation.
 \label{fig2}}
\end{figure}

\noindent 
\begin{figure} 
\caption{The angular dependence of the on-shell t-matrix as function 
of the laboratory energy from $E_{lab}=$ 50~MeV to $E_{lab}=$ 1000~MeV. 
On the left $Re T(q_0,q_0,x,E_{lab})$ is displayed, on the right
 $Im T(q_0,q_0,x,E_{lab})$. \label{fig3}}
\end{figure} 

\noindent 
\begin{figure} 
\caption{The angular dependence for the real part of the
half-shell t-matrices, $Re$
$T(q,q_0,x,E)$, is displayed for $E_{lab}=$ 200~MeV (a) and $E_{lab}=$
 500~MeV (b). \label{fig4}}
\end{figure} 

\noindent 
\begin{figure} 
\caption{The angular dependence for the real part of the
off-shell t-matrix, $Re$
$T(q,q'=250 MeV/c,x,E)$, is displayed for $E_{lab}=$ 400~MeV. \label{fig5}}
\end{figure} 

\noindent 
\begin{figure} 
\caption{The angular dependence for the real part of the 
off-shell t-matrix, $Re$
$T(q,q'=1000 MeV/c,x,E)$, is displayed for $E_{lab}=$ 400~MeV. \label{fig6}}
\end{figure} 

\noindent 
\begin{figure} 
\caption{The angular dependence for $(E-E_s)(E-E_p) T(q_0,q_0,x,E)$
for energies $E$ 
around the p-wave pole (a) and the s-wave pole (b). \label{fig7}}
\end{figure} 

\noindent 
\begin{figure} 
\caption{The angular dependence for $(E-E_s)(E-E_p) T(q_0,q_0,x,E)$
as function of the energy from E=-200~MeV to E=-1~MeV. Note the 
characteristic angular behavior around the p- and s-wave poles as well as
the strong forward peak between the two poles and below the s-wave
pole. \label{fig8}}
\end{figure}

\noindent 
\begin{figure} 
\caption{The angular dependence for the real part of the on-shell t-matrix
 $Re \; T(q_0,q_0,x,E)$ for $|E|$=200, 400 and 800~MeV. For comparison, the 
angular dependence of the driving term, $V^I$, is also shown as 
dash-dotted line.
 \label{fig9}}
\end{figure}

\noindent 
\begin{figure} 
\caption{The angular dependence for the real part of the
off-shell t-matrix $Re$
 $T(q,q',x,E=200 MeV)$ is displayed for q'=250 MeV/c (a) and for
q'=1000 MeV/c (b).  \label{fig10}}
\end{figure}

\noindent
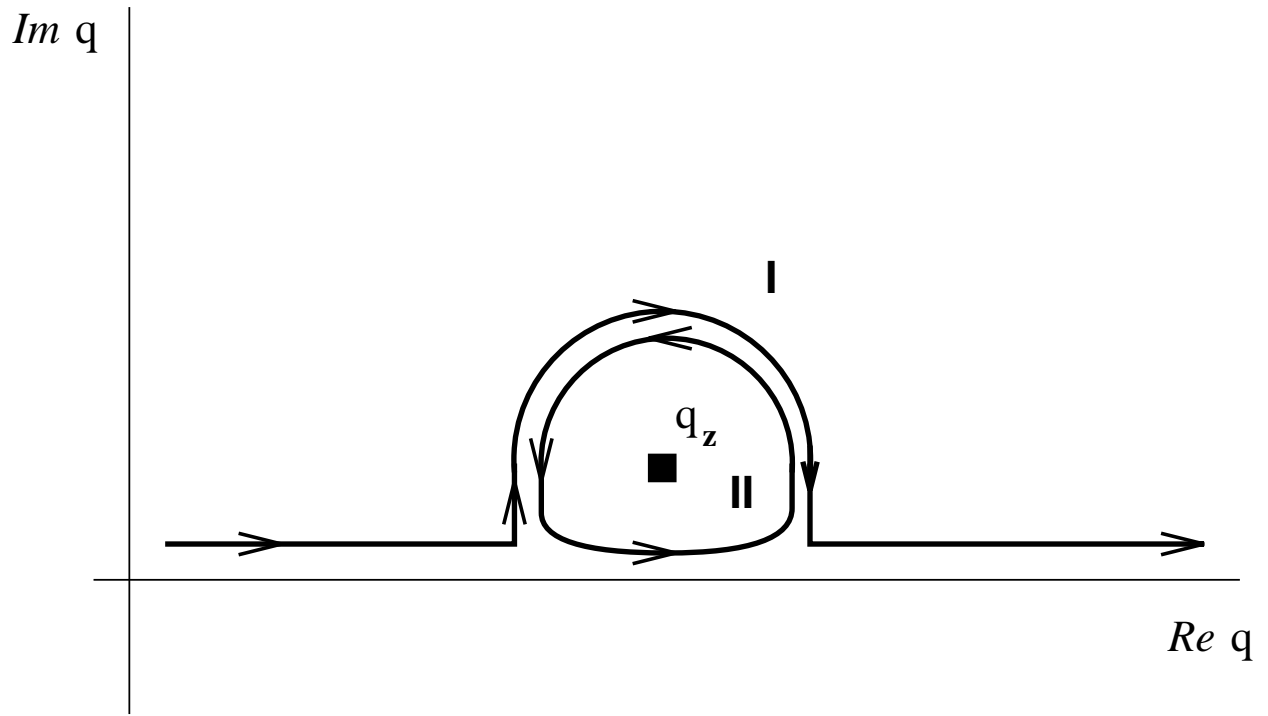
\begin{figure}
\caption{Modified integration path in the complex  $q$ plane for the
analytic continuation of the Lippmann-Schwinger equation into the
second energy sheet as described in the text. \label{fig11}}
\end{figure}

\noindent
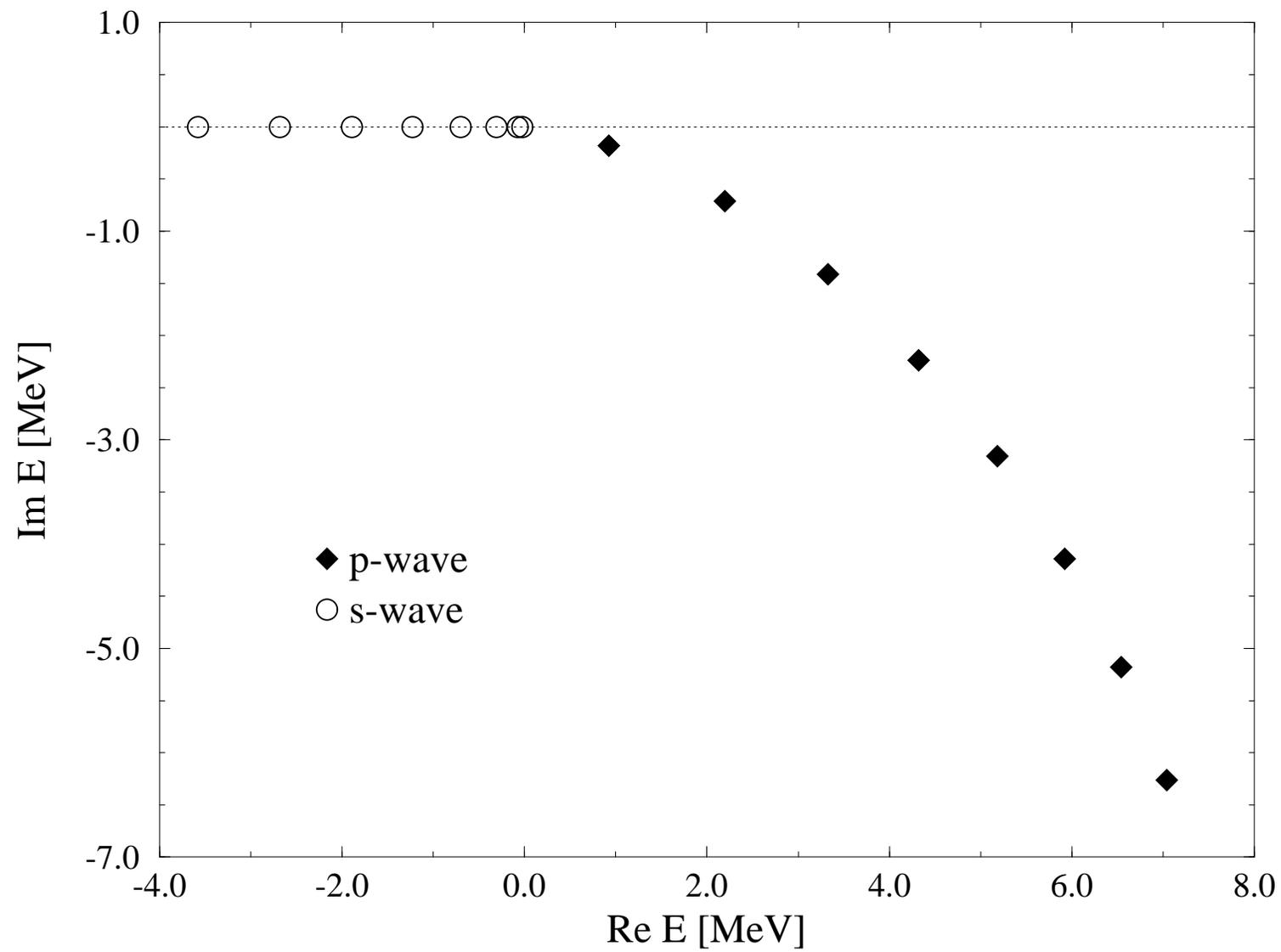
\begin{figure}
\caption{The pole trajectories of the p-wave and s-wave pole in
the second energy sheet as given by the potentials $V^{(IV)}$ and
$V^{(V)}$ of Table I. \label{fig12}}
\end{figure}

\noindent 
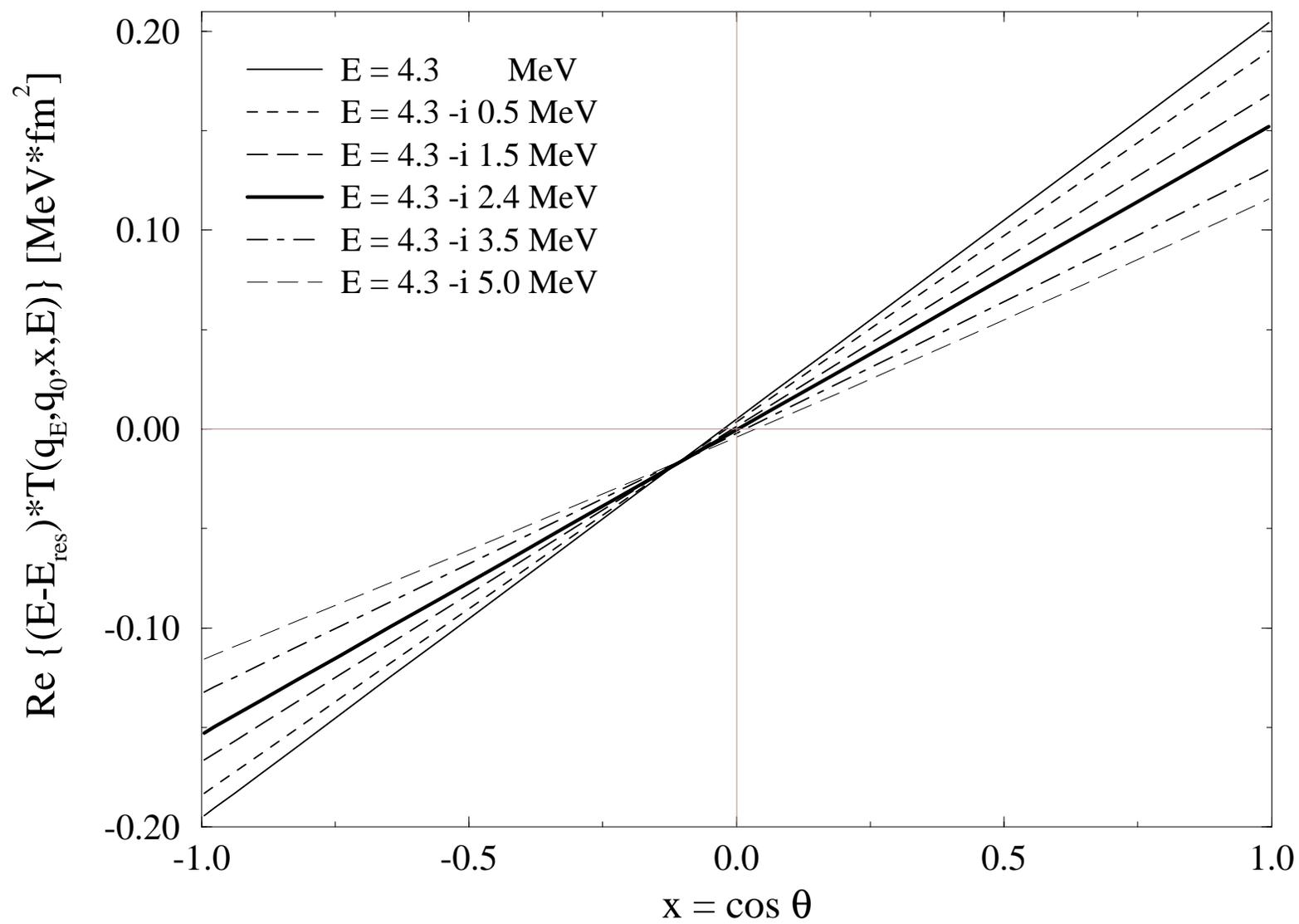
\begin{figure} 
\caption{The angular dependence for the real part of 
$(E-E_{res})T(q_E,q_0,x,E)$ as function of the complex energy E around
the p-wave resonance of potential model $V^{(IV)}$. The off-shell momentum
$q_0$ is fixed at 100~MeV/c.  \label{fig13}}
\end{figure}

\noindent 
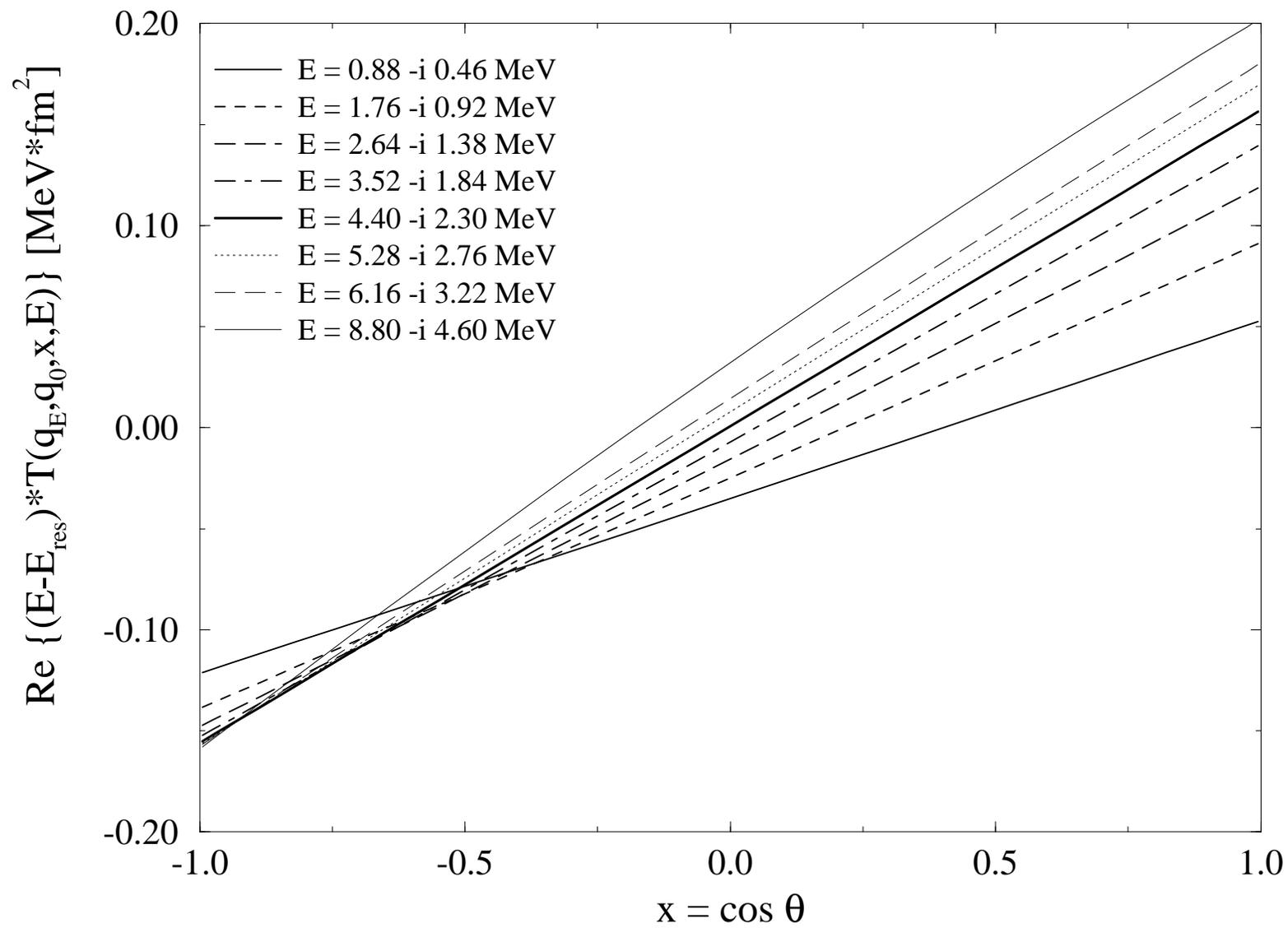
\begin{figure} 
\caption{Same as Fig.~13, but for a different path of the complex
energy E. \label{fig14}}
\end{figure}

\noindent 
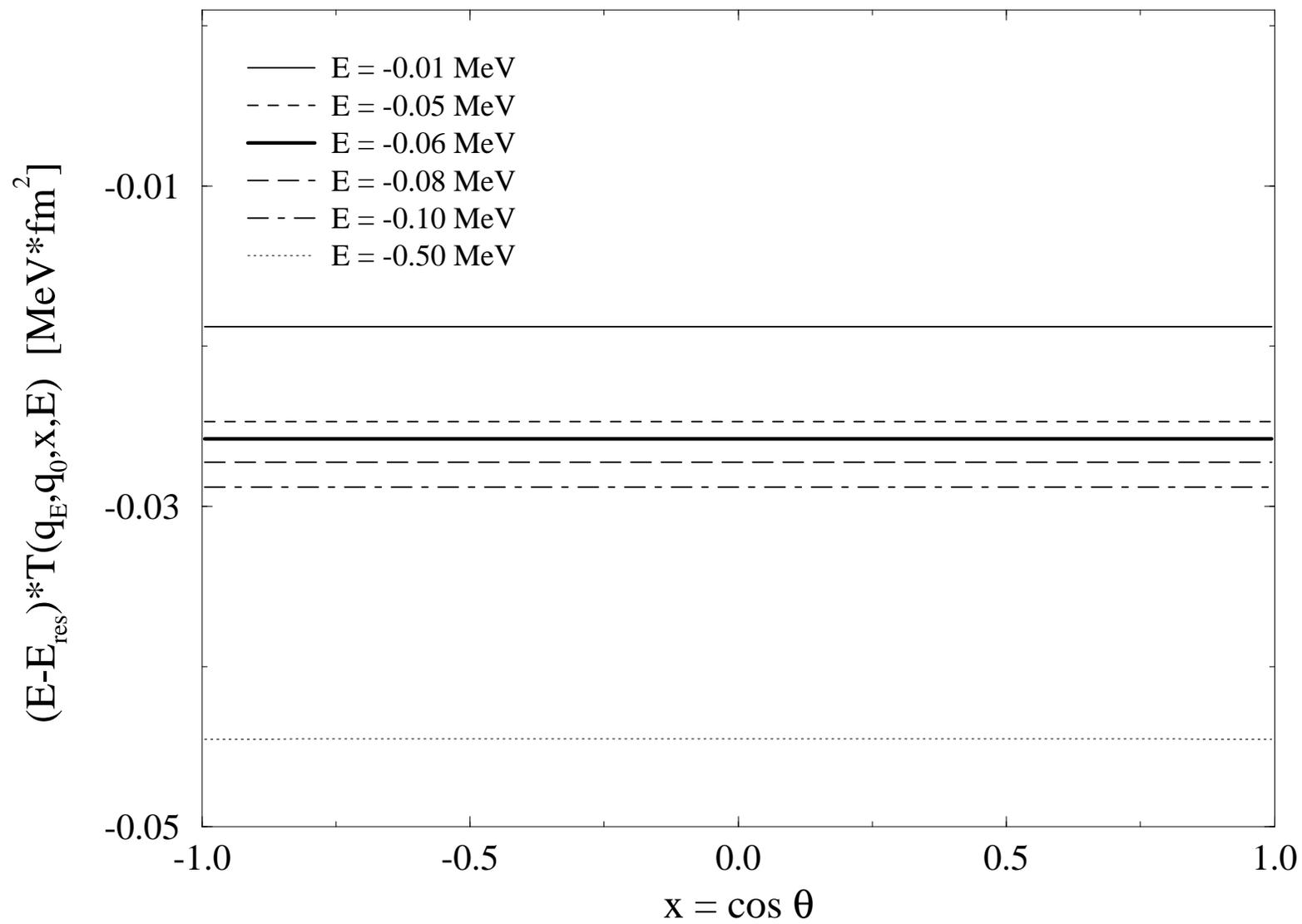
\begin{figure} 
\caption{The angular dependence of 
$(E-E_{res})T(q_E,q_0,x,E)$ as function of the negative energy E 
in the second energy sheet around
the s-wave virtual state of potential model $V^{(V)}$. The off-shell 
momentum $q_0$ is fixed at 100~MeV/c.  \label{fig15}}
\end{figure}


\begin{references} 
\bibitem{holz} J.~Holz and W.~Gl\"ockle,~Phys.~Rev {\bf C37}, 6 (1988);
J.~Holz and W.~Gl\"ockle, J. Comp. Phys. {\bf 76}, 131 (1988).
\bibitem{rice} R.A.~Rice and Y.E.~Kim, Few-Body Syst. {\bf 14}, 127
(1993).
\bibitem{hiltrop} J.~Hiltrop,~Diploma Thesis, Bochum, 1989, unpublished;
I.~Posukidis, Diploma Thesis, Bochum, 1993, unpublished.
\bibitem{huber} D. H\"uber, W.~Gl\"ockle, and A.~B\"omelburg, Phys. Rev 
{\bf C42}, 2342 (1990).

\bibitem{bonn} R.~Machleidt, K.~Holinde, Ch.~Elster, Phys. Rep. {\bf 149},
1 (1987).
\bibitem{redish} E.F.~Redish and K.~Stricker-Bauer, Phys. Rev. {\bf C36},
513 (1987).

\bibitem{MT} R.A.~Malfliet and J.A.~Tjon, Nucl. Phys. {\bf A127}, 161
 (1969).
\bibitem{ndprep} W.~Gl\"ockle, H.~Witala, D.~H\"uber, H.~Kamada, 
J.~Golak, Phys. Rep. {\bf 274}, 107 (1996).

\bibitem{Wbook} W.~Gl\"ockle, The Quantum Mechanical Few-Body Problem,
Springer Verlag, 1983.

\bibitem{weinberg} S.~Weinberg, Phys. Rev {\bf 133B}, 232 (1964).
 
\end{references}
\end{document}